\documentclass{article}
\usepackage{amsmath}
\usepackage{graphicx} 
\usepackage{xcolor,colortbl}
\usepackage{amssymb}
\usepackage{url}
\date{}

\title{Oscillator Formulations of Many NP Problems}
\author{Wenxiao Cai, Zongru Li, Yu-Neng Wang, Sara Achour,Thomas H. Lee \\ Stanford University}

\begin{document}

\maketitle

\footnotetext[1]{The authors are with the Department of Electrical Engineering, Stanford University, Stanford, CA 94305, USA. Correspond to: Thomas. H. Lee: tomlee@ee.stanford.edu, and Wenxiao Cai: wxcai@stanford.edu .}

\tableofcontents

\begin{abstract}
    Efficiently optimizing Nondeterministic Polynomial time (NP) problems in polynomial time has profound implications in many domains. 
    CMOS oscillator networks have been shown to be effective and efficient in approximating certain NP-hard problems such as minimization of Potts Hamiltonian, and computational complexity theory guarantees that any NP problem can be reduced to it. 
    In this paper, we formulate a variety of NP problems using first-order and multi-phase Potts Hamiltonian.
    We also propose a 3-state asymmetrically weighted oscillator optimizer design to optimize the problems. 
    Building on existing knowledge in CMOS design, our proposed algorithms offer a promising pathway for large-scale optimization of NP problems.    
\end{abstract}

\section{Introduction}
Optimizing NP problems in polynomial time is of great significance~\cite{np_1,np_2}. 
On conventional von Neumann architecture-based computers, finding the optimal solution requires exhaustively searching through all possible configurations. 
As problem complexity increases, the execution time typically grows super-polynomially.
Leveraging specially designed, non–von Neumann hardware architectures to approximate NP problems in polynomial time offers a promising alternative and holds substantial potential for advancing both theory and application.
Previous works~\cite{richelle_graph_coloring} place oscillators on CMOS circuits and encode information in their phases, enabling oscillator interactions to collectively minimize first-order and multi-phase Potts Hamiltonian, which is a famous NP-hard problem.
Large-scale CMOS technologies~\cite{cmos_1,cmos_2} enable the deployment of large numbers of oscillators in hardware, which in turn enables us to optimize more complex problems. 
In theory, every NP problem can be reduced to oscillator formulations but the reduction is unexplored.
Among them, Karp's 21 NP-complete problems~\cite{karp} are the earliest and most well-known examples.
In this paper, we efficiently reduce 19 of Karp's 21 NP-complete problems and many other P problems to oscillator formulations and propose the design of 3-state asymmetrically weighted oscillator optimizer. 
Leveraging well-established fabrication processes and the rich dynamical behavior of oscillators~\cite{cmos_oscillator_2}, this approach offer a practical and scalable computational paradigm.

\subsection{Difference Between Ising and Oscillator Formulation of Potts Hamiltonian}

Our multi-phase oscillator networks~\cite{richelle_graph_coloring,oscnet} encode information in their phases and minimize Potts Hamiltonian.
The possible phase $\theta$ is $n \frac{2\pi}{N}$, where $N$ is the integer multiple relationship between the main pump and the intrinsic frequency of the oscillator, and $n=0, 1,...,N-1$.
In our multi-phase CMOS oscillator system, we choose $N=3$.

While Ising machines~\cite{ising_many_np} 
and our multi-phase oscillators can both minimize Potts Hamiltonian, there as still some difference between them:
\begin{itemize}
    \item Instead of only encoding the solution in the global minimum of Potts Hamiltonian~\cite{oim,ising_1,ising_2}, we introduce an additional phase to account for the dynamics of phase updating and reduce the number of connections.
    \item Different from Ising based 3-SAT optimizers~\cite{oim,dikopoulos202525}, we directly formulate many NP problems with the oscillator optimizer and use mixed k-SAT if needed, both significantly reducing the number of nodes and connections.
    \item Formulations in Ising machines~\cite{ising_many_np} and Quantum Computing~\cite{quantum_np_1} is Quadratic Unconstrained Binary Optimization (QUBO), while ours belong to Linear Unconstrained Multi-valued Optimization.

\end{itemize}

\section{Background}
\subsection{Analog Computing on Multi-phase Oscillators}
Multi-phase CMOS Oscillator Networks~\cite{richelle_graph_coloring,richelle_quantum,oscnet} operates with high-order injection locking, with local connections to neighboring oscillators. All oscillators are injected with a master pump signal at $N·f_0$, where $f_0$ is oscillators' natural frequency, and $N$ is a chosen integral.
The phases of oscillators can be $n \frac{2\pi}{N}$, where $n=1,2...N-1$.
This network mimics the activity of polychronous spiking neurons in the brain, where each oscillator represents a single brain cell. 
Information is encoded in the phases of the individual oscillators. 
The voltage of oscillator $i$ at time $t$ is:
\begin{equation}
    V_i(t)=A_i cos(\omega_0t+\theta_i),
\end{equation}
where $\omega_0$ is the free-running angular frequency of the oscillator, $A_i$ is the amplitude, and $\theta_i$ is the phase.
Each oscillator outputs a current based on its own state, while changes in its phase are determined by the influence of neighboring oscillators and the main pump, which can be explained by Impulse Sensitivity Function (ISF) theory~\cite{isf_1}.
Phase changes of oscillator $i$, connected to neighboring oscillators, can be represented by Kuramoto’s equation~\cite{kuramoto_eq}:
\begin{equation}
    \frac{d\theta_i}{dt} = \sum_{(i,j) \in \mathcal{N}} K_{ij} sin(\theta_i - \theta_j) + K_p sin(N\theta_i),
    \label{eq:kuramoto}
\end{equation}
where $k_{i,j}$ represents the coupling strength between oscillator $i$ and $j$,  $K_p$ models the main pump current and ISF, and $\mathcal{N}$ is the set of neighboring oscillators of $i$.
The global Lyapunov function exists and will be minimized over time by the oscillator network~\cite{richelle_graph_coloring}: 
\begin{equation}
    E(t) = \frac{N}{2} \sum_{(i,j) \in  \mathcal{N}} K_{ij} cos(\theta_i - \theta_j) + \sum_{i} K_p cos(N\theta_i).
    \label{eq:Lyapunov}
\end{equation}

Eq.~\ref{eq:Lyapunov} takes a similar form of Potts Hamiltonian:
\begin{equation}
    H = - \sum_{(i,j) \in \mathcal{N}} J_{ij} cos(\theta_i - \theta_j) - \sum_{i} h_i.
    \label{eq:potts}
\end{equation}

The CMOS oscillator network can optimize Potts Hamiltonian, which is a famous NP-hard problem. 
In this paper, we propose to reduce many NP problems to Potts Hamiltonian. 
We then propose an oscillator optimizer to minimize Potts Hamiltonian~\cite{richelle_graph_coloring}, thus optimizing a bunch of NP problems.

\subsection{3-State Potts Hopfield Model}
In this paper, we focus on Potts Hopfield Model~\cite{hopfield,potts_hopfield} with 3 phases.
In our 3-state model, each neuron $i$ can be in one of three states, $\sigma_i \in \{F, T, B\}$, which stands for True, False and Blue. Without loss of generality, let $\{F, T, B\}$ correspond to $\{0,\frac{2}{3}\pi,\frac{4}{3}\pi \}$ in the oscillator network.
It is convenient to represent these discrete states as two-dimensional vectors, $\vec{S}_i \in \{\vec{s}_F, \vec{s}_T, \vec{s}_B\}$, pointing to the vertices of an equilateral triangle centered at the origin. These vectors can be defined as:
\begin{align}
    \vec{s}_F &= \begin{pmatrix} 1 \\ 0 \end{pmatrix} \\
    \vec{s}_T &= \begin{pmatrix} \cos(2\pi/3) \\ \sin(2\pi/3) \end{pmatrix} = \begin{pmatrix} -1/2 \\ \sqrt{3}/2 \end{pmatrix} \\
    \vec{s}_B &= \begin{pmatrix} \cos(4\pi/3) \\ \sin(4\pi/3) \end{pmatrix} = \begin{pmatrix} -1/2 \\ -\sqrt{3}/2 \end{pmatrix}
\end{align}
The interaction between any two states $\alpha, \beta \in \{F, T, B\}$ in neighboring node pair $(i,j)$ is naturally given by the dot product of their corresponding vectors, $J_{ij} \vec{s}_\alpha \cdot \vec{s}_\beta$. This defines an interaction matrix $W^{\alpha\beta}$:
\begin{equation}
    W_{ij}^{\alpha\beta} = J_{ij} \vec{s}_\alpha \cdot \vec{s}_\beta = 
    J_{ij}
    \begin{pmatrix}
        1 & -1/2 & -1/2 \\
        -1/2 & 1 & -1/2 \\
        -1/2 & -1/2 & 1
    \end{pmatrix}
\end{equation}

We define the local field vector $\vec{h}_i$ acting on neuron $i$:
\begin{equation}
    \vec{h}_i = \sum_{j \ne i} \mathbf{J}_{ij} \vec{S}_j = \frac{1}{N} \sum_{j \ne i} \sum_{\mu=1}^{p} \vec{\xi}_i^\mu (\vec{\xi}_j^\mu \cdot \vec{S}_j)
\end{equation}

\begin{figure}[h]
	\begin{center}
    \includegraphics[width=0.8\linewidth]{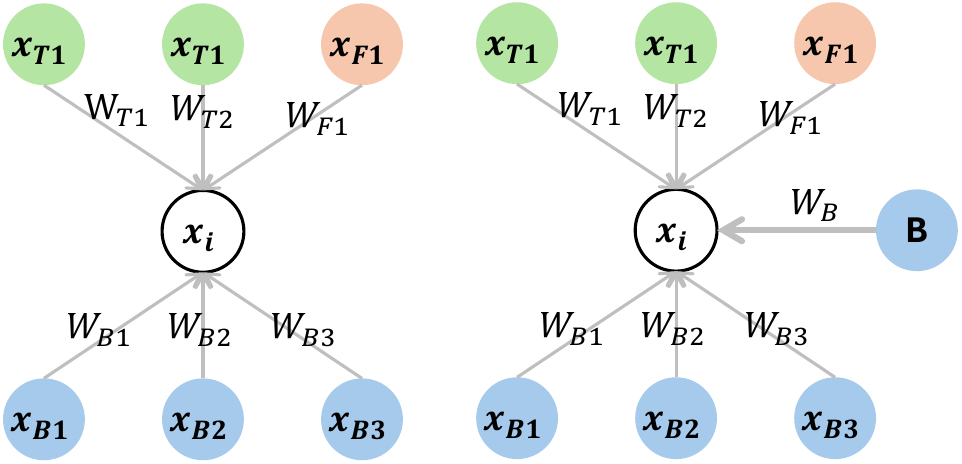}
	\end{center}
\caption{Left: updating $x_i$ in 3-state Potts Hopfield Network. Right: a special case when $x_i$ is connected to Blue with a large positive weights. }
	\label{fig:3_state_potts_hopfield_update}
\end{figure}

A neuron $i$ updates its state to cause the largest decrease in energy in Eq.~\ref{eq:Lyapunov}. The state of neuron $i$ at the next time step, $\vec{S}_i(t+1)$, is chosen from the set of possible state vectors $\{\vec{s}_F, \vec{s}_T, \vec{s}_B\}$ according to:
\begin{equation}
    \vec{S}_i(t+1) = \underset{\vec{s} \in \{\vec{s}_1, \vec{s}_2, \vec{s}_3\}}{\text{argmin}} (\vec{s} \cdot \vec{h}_i(t))
    \label{eq:potts_hopfield_update}
\end{equation}
This update is performed until the network settles into a stable state, which corresponds to a local minimum of the Hamiltonian. 
An example is given in Fig.~\ref{fig:3_state_potts_hopfield_update}. To update the phase of $x_i$, we just add up the weights of each color connected to it. In Potts Hamiltonian, if a weight is positive, then we would like the pair of nodes to choose different colors. If a weight is negative, then the nodes tend to converge to the same color. So, in this example, if $W_{T1} + W_{T2} < W_{F1}$ and $W_{T1} + W_{T2} < W_{B1} + W_{B2} + W_{B3}$ , then $x_i$ is updated to True color.
When $x_i$ is already coupled with voltage source Blue with a very large weight $W_B >> W_{T1} + W_{T2}$ and $W_B >> W_{F1}$ , then $x_i$ will only choose between True and False. Any Blue node connected to $x_i$ have no preference over $x_i$ choosing True or False. This is simple rule but we will use it a lot in designing optimizers.

\section{Oscillator Optimizer Network}
In this section, we show the design of our 3-state oscillator optimizer for SAT problem. In designing, our oscillator optimizer takes dynamics of phase update into consideration. It is based on analog computing only so it is much faster and more efficient than oscillator optimizers with digital components~\cite{dikopoulos202525}. 

\subsection{Network Architecture and Base Network}
\begin{figure}[h]
	\begin{center}
    \includegraphics[width=0.8\linewidth]{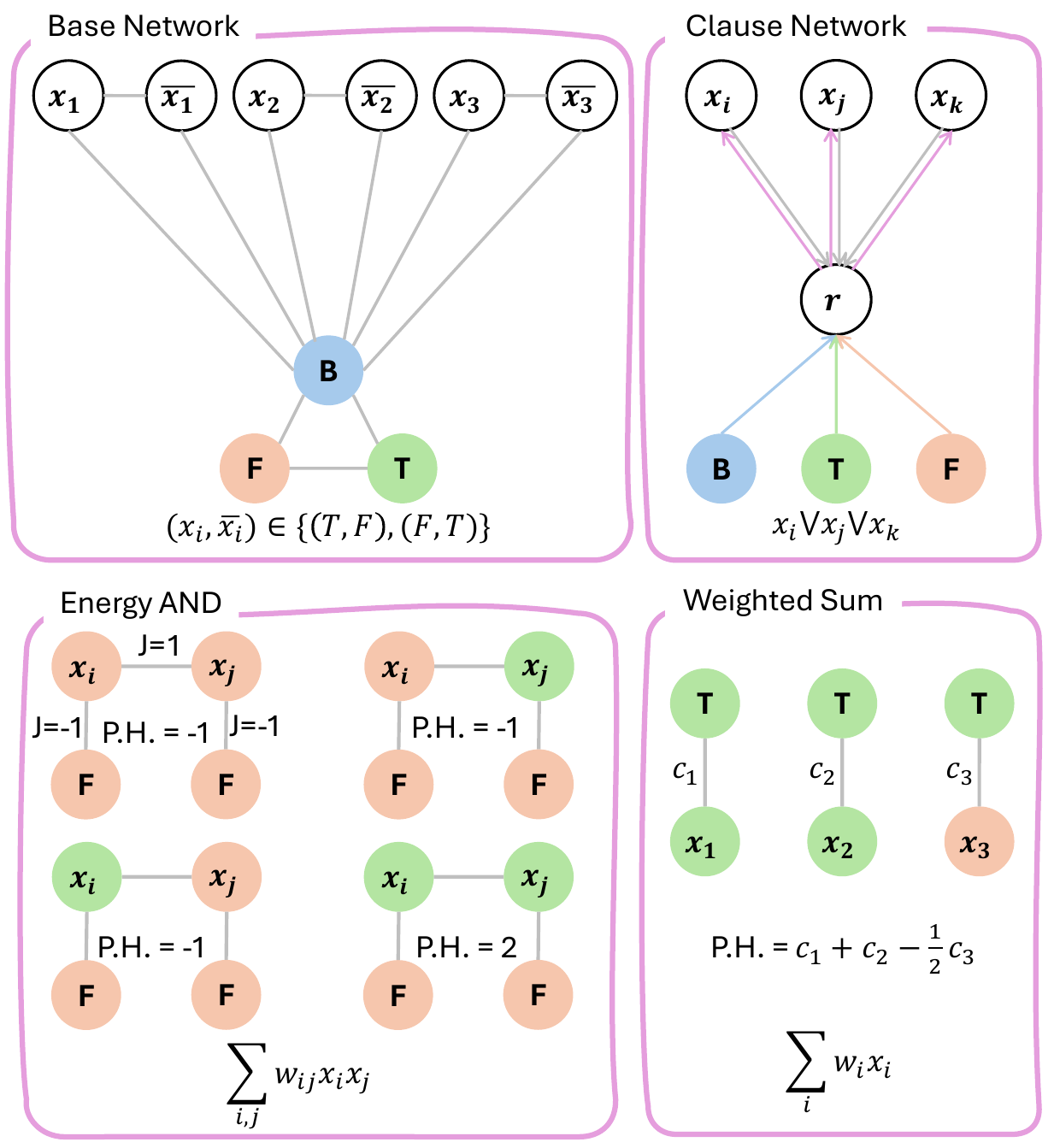}
	\end{center}
    \vspace{-1em}
\caption{Our optimizer is composed of four parts: base network, clause network, Energy AND network, and Weighted Sum network. Blue, True and False are voltage sources with fixed phases.}
	\label{fig:optimizer_overall}
\end{figure}

Our optimizer is composed of four parts: base network, clause
network, Weighted Sum network, and Energy AND network, as shown in Fig.~\ref{fig:optimizer_overall}. 
B, T, F stands for Blue, True and False and are fixed at state $\vec{s}_T, \vec{s}_F$ and $\vec{s}_B$.
In base network, $x_i$, $\bar{x_i}$ and B are connected to each other, so $(x_i, \bar{x_i}) \in \{(T,F),(F,T)\}$.
The variables are connected to a clause network for each clause. $l_1$ to $l_3$ in each clause network represent the variables in each SAT clause.

\subsection{Dynamics of Phase Update in Clause Network}
\begin{figure}[h]
	\begin{center}
    \includegraphics[width=0.5 \linewidth]{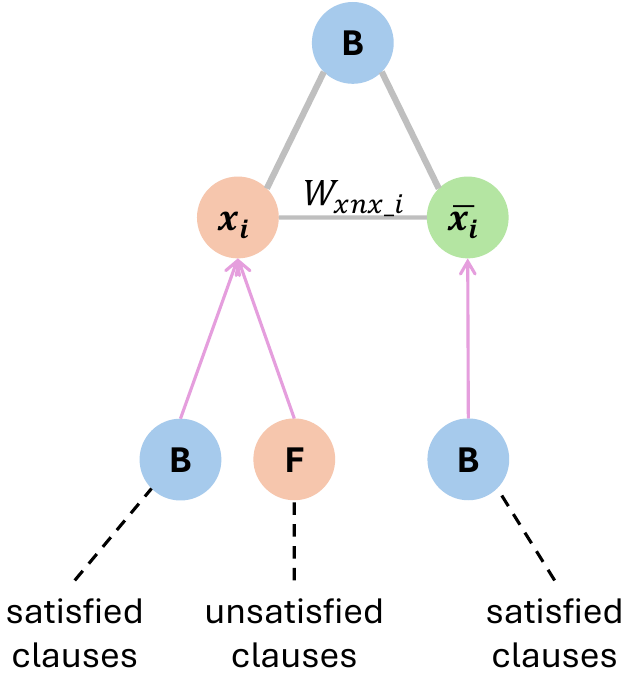}
	\end{center}
\caption{ Dynamics of phase updating in our oscillator optimizer.}
	\label{fig:dynamics_phase_update}
\end{figure}

Without loss of generality, consider $x_i=\text{False}$ and $\bar{x_i}=\text{True}$ in an assignment of SAT.
The update dynamics is shown in Fig.~\ref{fig:dynamics_phase_update}. Only the unsatisfied clauses that variable $x_i$ is in give feedback to $x_i$ and motivate it to change from False to True. The satisfied clauses are connected to $x_i=\text{True}$ and $\bar{x_i}=\text{True}$ with Blue, which do not have any preference over $x_i$ and $\bar{x_i}$ choosing True or False.
Overall, the unsatisfied clauses will push the system to change the phases of corresponding oscillators. 

\subsection{Connection Weights of Clause Network}

\begin{figure}[h]
	\begin{center}
    \includegraphics[width=1\linewidth]{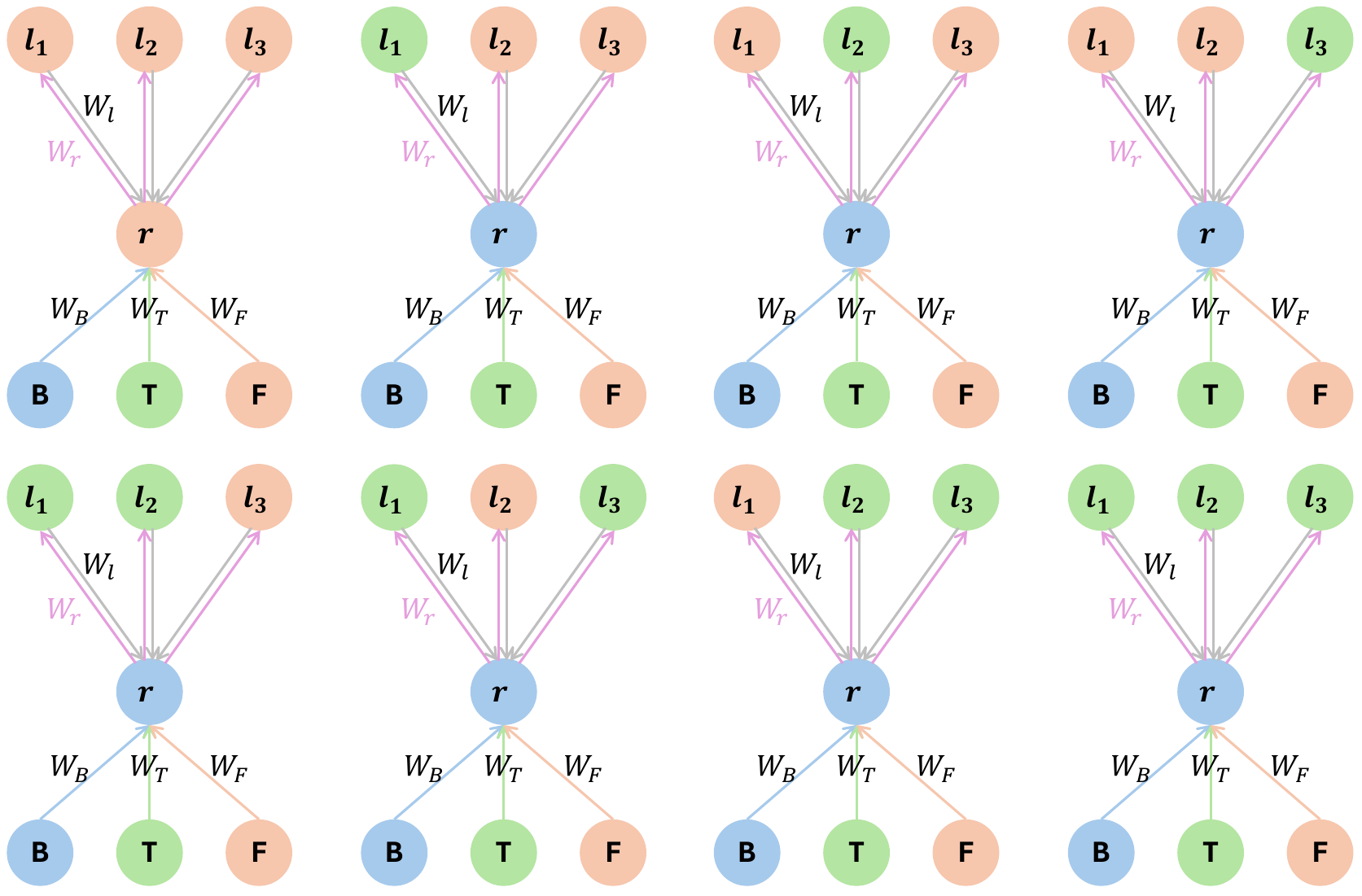}
	\end{center}
\caption{Truth table of our optimizer design. $l_1$ to $l_3$, together with three voltage sources True, False and Blue, decide the phase of additional variable $r$. $r$ provides feedback that influences $l_1$ to $l_3$. The weights are asymmetrical: $W_l \neq W_r$.}
	\label{fig:optimizer_truth_table}
\end{figure}

In each clause network, our intuitive is that, when $x_i$ or $\bar{x_i}$ is in a unsatisfied clause, it should have a tendency to jump from False to True, and the other variable should change accordingly.
This leads to a truth table in Fig.~\ref{fig:optimizer_truth_table}. When all three variables $l_1$ to $l_3$ are all False and the clause is unsatisfied, we set the additional node $r=\text{False}$ and let it inform $l_1$ to $l_3$ to consider changing to True. Otherwise, when at least on of  $l_1$ to $l_3$ is True, the clause is satisfied and $r$ is set to Blue. $r=\text{Blue}$ do not have any preference over $l$ choosing True or False. This heuristic is that: since the clause is already satisfied, $l$ do not need to switch phase and can stay where it is now.
Following update rule in Eq.~\ref{eq:potts_hopfield_update} and truth table in Fig.~\ref{fig:optimizer_truth_table}, we use $l_1$ to $l_3$ to decide the phase of $r$:
\begin{equation}
    \begin{cases}
        &3W_l + W_F < W_B, \quad
        3W_l + W_F < W_T, \quad
        W_B < W_l + W_T \\
        &W_B < 2W_l + W_F, \quad
        W_B < 2W_l + W_T, \quad
        W_B < W_l + W_F \\
        &W_B < 3W_l + W_T, \quad
        W_B < W_F \\
    \end{cases}
    \label{eq:inequality_truth_table}
\end{equation}

In addition, we set $W_r$ to be positive so when $l_1=l_2=l_3=r=\text{False}$, $r$ will drive $l_1$,$l_2$,$l_3$ to True. The weights are asymmetrical because $W_l \neq W_r$. 
Together with Eq.~\ref{eq:inequality_truth_table}, the inequality can be simplified to:
\begin{equation}
    \begin{cases}
        &W_r > 0, \quad
        W_l < 0 \\
        &W_F + 3W_l < W_B < W_F \\
        &W_B < W_T + 3W_l
    \end{cases}
\end{equation}

The three voltage sources connected to $r$ can be merged in to one Biased Source ($\vec{s}_{BS}$).
\begin{align}
    \theta_{BS} &= -\arctan(\frac{\sqrt{3}(W_B-W_T)}{2W_F-W_B-W_T}) \\
    |\vec{s}_{BS}| &= \sqrt{W_F^2 + W_T^2 + W_B^2 - W_FW_T - W_FW_B - W_TW_B} 
\end{align}

\subsection{Energy AND Network}
We introduce Energy AND network to minimize the objective function $\sum_{i,j}w_{ij}x_i x_j$, where $x\in\{0,1\}$. 
The Energy AND gate leverages oscillator-based energy minimization to emulate logical conjunction (AND).
We introduce additional oscillators fixed at False phase. In the Energy AND network showin in the bottom left part of Fig.~\ref{fig:optimizer_overall}, only when $x_i=x_j=T$, the Potts Hamiltonian is 2. Otherwise, it is -1. The Potts Hamiltonian $\sum_{i,j}w_{i,j}(cos(\theta_i-\theta_j)+cos\theta_i+cos\theta_j)=\sum_{i,j}w_{i,j}(3x_ix_j-1)$is linear to the objective function $\sum_{i,j}w_{i,j}x_ix_j$. 
Minimizing Potts Hamiltonian will be effectively same as minimizing the desired objective function with AND logic between $x_i$ and $x_j$.

\subsection{Weighted Sum Network}
\label{sec:weighted_sum_network}
In this part of the network, we would like to minimize the objective function: $\sum_{i} c_i x_i$, where $x_i \in \{0,1\}$.
To do so, we connect each $x_i$ with a oscillator fixed at True phase, with coupling strength $c_i$. An example is shown in the bottom right of Fig.~\ref{fig:optimizer_overall}.
Since $x_i$ is constrained to be True or False, the Potts Hamiltonian is $\sum_{i} c_i cos(\theta_i-\frac{2}{3}\pi) = \sum_{i} c_i z_i$, where $z_i \in \{-0.5,1\}$. 
The solution of $x$ in oscillator in Potts Hamiltonian is exactly the same as that of original objective function, as linear transformation $x_i = \frac{2}{3}(z_i+0.5)$ will not affect the solution.

\section{Many NP Problems Formulated by Oscillator Optimizer}
We formulate 19 of Karp's 21 NP-complete problems~\cite{karp} and some of their extensions with the oscillator optimizer.
For some problems, we formulate the NP-hard optimization version.
SAT and 3-SAT have been discussed in previous sections.
We also analyze the number of variables, the number of clauses, and the clause lengths if a problem, or part of the problem, is directly reduced to SAT.
The number of oscillators and connections of SAT is discussed in Sec.~\ref{sec:osc_connect_sat}.
2 of Karp’s 21 NP-complete problems can be formulated for the oscillator optimizer. However, we do not discuss them here because, to the best of our knowledge, their formulations require tree-like structures that make the oscillator approach inefficient. Specifically, Sequencing requires an addition circuit with a tree topology, and Steiner Tree similarly relies on a tree structure.

\subsection{0-1 Integer Programming and Optimization}
\label{sec:IP}

Consider the following 0-1 Integer Programming (IP) problem:
\begin{equation}
\begin{aligned}
\text{Given:} \quad & C \in \mathbb{Z}^{m \times n},\quad b \in \mathbb{Z}^m \\
\text{Find:} \quad & x \in \{0,1\}^n \quad \text{such that } C x = b
\end{aligned}
\label{eq:ip_definition}
\end{equation}

The problem can also be written as:
\begin{equation} 
    \sum_{j=1}^{n} c_{ij}x_j = b_i, \text{for} \quad i=1,\text{...},m
    \label{eq:ip}
\end{equation}

Eq.~\ref{eq:ip} can be transformed to energy minimization:
\begin{equation} 
\begin{aligned}
        \text{min} \quad &\sum_{i=1}^{m} (\sum_{j=1}^{n} c_{ij}x_j - b_i)^2 \\
        =& \sum_{i=1}^{m} (\sum_{j=1}^{n} c_{ij}^2x_j^2 + b_i^2 + 2\sum_{1\leq j<k\leq n}c_{ij} c_{ik} x_j x_k - 2b_i\sum_{j=1}^{n} c_{ij} x_j ) \\
        =& \sum_{j=1}^{n} (\sum_{i=1}^{m} c_{ij}^2) x_j^2 + \sum_{1\leq j<k\leq n} (\sum_{i=1}^{m}c_{ij} c_{ik}) x_j x_k - \sum_{j=1}^{n}(\sum_{i=1}^{m}2b_i c_{ij}) x_j
\end{aligned}
\end{equation}

Since $x\in\{0,1\}$, $x^2=x$, the energy minimization term can be written as:
\begin{equation} 
\begin{aligned}
    \text{min} \quad \sum_{i=1}^{m} \sum_{j=1}^{n} (c_{ij}^2 - 2b_i c_{ij}) x_j + \sum_{1\leq j<k\leq n} \sum_{i=1}^{m} 2c_{ij} c_{ik} x_j x_k 
\end{aligned}
\end{equation}

In our oscillator optimizer, the $x_j$ terms are mapped to Weighted Sum network, and $x_j x_k$ terms can be converted to Energy AND network.
For the 0-1 Integer Optimization problem:
\begin{equation}
\begin{aligned} \label{0-1io}
\text{min} \quad & a^\top x \\
\text{s.t.} \quad & C x = b \\
& x \in \{0,1\}^n
\end{aligned}
\end{equation}
To minimize the objective function, we use add $\lambda a^\top x$ to the Weighted Sum network as discussed in the oscillator network of 0-1 IP in Eq.~\ref{eq:ip_definition}. 
We set $0<\lambda<1$ so the constraints are first satisfied.
The overall Potts Hamiltonian is:
\begin{equation} 
\begin{aligned}
    \text{min} \quad (\lambda a^\top + \sum_{i=1}^{m} \sum_{j=1}^{n} (c_{ij}^2 - 2b_i c_{ij})) x_j + \sum_{1\leq j<k\leq n} \sum_{i=1}^{m} 2c_{ij} c_{ik} x_j x_k 
\end{aligned}
\end{equation}

For both 0-1 Integer Programming and Integer Optimization, The number of oscillators is $n$, and the number of connections is $n+\frac{3}{2}n(n-1)$. 
Compared to reducing Integer Programming to SAT with binary add circuits and optimizing the problem with SAT optimizer~\cite{een2006translating}, which requires $O\!\left(\sum_{i=1}^m n_i \log S_i \right)$ nodes and connections ($S_i = \sum_{j=1}^n |c_{ij}|$), 
our formulation saves the number of nodes and connections by orders of magnitude in general cases.

\subsection{Directed and Undirected Hamilton Path and Circle}
Directed and Undirected Hamilton Circle count as two in Karp's 21 NP-complete problems~\cite{karp}. We reduce both directed and undirected Hamiltonian Circle and Hamilton Path with $n$ nodes to n-SAT.
The problems are defined as follows:

\begin{quote}
    Given a directed or undirected graph $G = (V, E)$, where $V = \{v_1, v_2, \dots, v_n\}$ is the set of vertices and $E$ is the set of edges, determine whether there exists a cycle that visits each vertex exactly once and returns to the starting vertex.
\end{quote}

We define boolean variables $X_{i,j}$ for each vertex $v_i \in V$ and position $j$ in the sequence (where $1 \leq j \leq n$). Here, $X_{i,j} = 1$ if and only if vertex $v_i$ appears in position $j$ in the Hamiltonian circuit. Otherwise,  $X_{i,j} = 0$. We can reduce the problems to n-SAT by introducing the $6$ constraints:

\textbf{Constraint I}: each position in the sequence must be occupied by exactly one vertex:
\begin{equation}
    \bigvee_{i=1}^{n} X_{i,j}, \quad \forall j \in \{1, \dots, n\}
\end{equation}

\textbf{Constraint II}: no two vertices occupy the same position:
\begin{equation}
    \neg (X_{i,j} \wedge X_{k,j}) = (\neg X_{i,j} \bigvee \neg X_{k,j}), \quad \forall i \neq k, \forall j
\end{equation}

\textbf{Constraint III}:
each vertex must appear exactly once in the cycle:
\begin{equation}
    \bigvee_{j=1}^{n} X_{i,j}, \quad \forall i \in \{1, \dots, n\}
\end{equation}

\textbf{Constraint IV}: a vertex cannot appear in multiple positions:
\begin{equation}
    \neg (X_{i,j} \wedge X_{i,k}) = (\neg X_{i,j} \bigvee \neg X_{i,k}), \quad \forall j \neq k, \forall i
\end{equation}

\textbf{Constraint V}:
A valid Hamiltonian circuit must follow the given graph edges. If vertex $v_i$ appears at position $j$ and vertex $v_k$ appears at position $j+1$, then there must be an edge between $v_i$ and $v_k$ in $E$:
\begin{equation}
    X_{i,j} \wedge X_{k,j+1} \Rightarrow (i,k) \in E
\end{equation}
This can be rewritten in CNF form:
\begin{equation}
    \neg X_{i,j} \vee X_{k,j+1} \vee X_{m,j+1} \vee \dots
\end{equation}
where $k, m, \dots$ is the set of nodes that are adjacent to node $i$: $(i,k) \in E$, $(i,m) \in E \dots$

\textbf{Constraint VI}: for Hamilton Circle only,
the last vertex in the sequence must connect back to the first vertex:
\begin{equation}
    \neg X_{i,n} \vee X_{k,1} \vee X_{m,1} \vee \dots
\end{equation}
where $k, j, \dots$ is the set of nodes that are adjacent to node $i$: $(i,k) \in E$, $(i,m) \in E \dots$.

Let $N_i$ denote the number of neighbors of $v_i$ in $E$, and let $|E|$ denote the total number of edges, so $\sum_{i=1}^nN_i=|E|$. 
he number of variables and connected added to reduce k-SAT to 3-SAT in Constraint V and VI is: $n\sum_{i=1}^n \min (N_i,1)$. We assume that there should be at least one solution of the input problem, so $N_i\geq1$, thus $n\sum_{i=1}^nN_i = 2|E|n$.
The number of variables, clauses, average and total length of mixed k-SAT and 3-SAT optimizer with our formulation of Hamilton Path and Circle is shown in Table.~\ref{tab:v_c_hamilton_path} and Table.~\ref{tab:v_c_hamilton_circle} respectively.
It's notable that mixed k-SAT optimizer saves an order of magnitude in the number of clauses and a huge amount of variables compared to a 3-SAT optimizer. 

\begin{table}[h]
\centering
\caption{Number of variables ($\#v$), clauses ($\#c$), average ($l_{avg}$) and total ($l_{total}$) length of mixed k-SAT and 3-SAT optimizer with our formulation of Hamilton Path.}
\setlength{\tabcolsep}{8pt} 
\renewcommand{\arraystretch}{1.5} 
\begin{tabular}{c|cc}
\hline
Optimizer & Mixed k-SAT & 3-SAT\\
\hline
$\#v$ & $n^3+2n$ & $2n^3+n^2-4n+n\sum_{i=1}^{n-1}N_i$\\
$\#c$ & $n^2$ & $n^3+2n^2-6n+n\sum_{i=1}^{n-1}N_i$ \\
$l_{avg}$ & $\frac{2n^2+\sum_{i=1}^{n-1}N_i}{n^2+2}$& $3$ \\
$l_{total}$ & $2n^3+\sum_{i=1}^{n-1}N_i$ & $3(2n^3+n^2-4n+n\sum_{i=1}^{n-1}N_i)$ \\
\hline
\end{tabular}
\label{tab:v_c_hamilton_path}
\end{table}

\begin{table}[h]
\centering
\caption{Number of variables ($\#v$), clauses ($\#c$), average ($l_{avg}$) and total ($l_{total}$) length of mixed k-SAT and 3-SAT optimizer with our formulation of Hamilton Circle.}
\setlength{\tabcolsep}{8pt} 
\renewcommand{\arraystretch}{1.5} 
\begin{tabular}{c|cc}
\hline
Optimizer & Mixed k-SAT & 3-SAT\\
\hline
$\#v$ & $n^3+2n$ & $2n^3+n^2-4n+2|E|n$\\
$\#c$ & $n^2$ & $n^3+2n^2-6n+2|E|n$ \\
$l_{avg}$ & $\frac{2n^2+2|E|}{n^2+2}$& $3$ \\
$l_{total}$ & $2n^3+2|E|n$ & $6n^3+3n^2-12n+6|E|n$ \\
\hline
\end{tabular}
\label{tab:v_c_hamilton_circle}
\end{table}

\subsection{Traveling Salesman Problem}
\label{sec:TSP}

The Traveling Salesman Problem (TSP) is defined as follows:
\begin{quote}
    Given a complete weighted graph $G = (V, E)$, where $V = \{v_1, v_2, \dots, v_n\}$ is the set of cities and $W$ assigns a cost $W_{ij}$ to travel from city $v_i$ to city $v_j$, determine whether there exists a cycle that visits each city exactly once and returns to the starting city with total cost at most $K$, for a given constant $K$.
\end{quote}

The variables of TSP is defined the same as Hamilton Circle.
Suppose there are $n$ cities in TSP, the available edges are $E$, with the weight, or traveling length $W_{uv}$ between city $u$ and city $v$.
Oscillator optimizer design for TSP is composed of two parts. First, a Clause network for Hamilton Circle. The weight of each connection in SAT is larger or equal to $\sum_{(uv)\in E} W_{uv}$ to ensure that the engine always find a valid Hamilton Circle as its candidate solution.
Second, we want to make the total traveling distance small by using the Energy AND network. 
In TSP, $n^2$ oscillators are connected to their neighbors and the Potts Hamiltonian of this part is:
$\sum_{(uv)\in E} W_{uv} \sum_{j = 1}^{n} (3x_{u,j}x_{v,j+1} - 1)$. 
Since this energy function is linearly transformed from
$\sum_{(uv)\in E} W_{uv} \sum_{j = 1}^{n} x_{u,j}x_{v,j+1}$, $x$ derived from them should be the same.
This second part of oscillators are all connected to the Blue oscillator in Graph Coloring so they can only converge to phases of True and False.



\subsection{Clique}

The Clique problem is defined as follows:
\begin{quote}
    Given an undirected graph $G = (V, E)$ and an integer $k$, determine whether there exists a subset $C \subseteq V$ of size $|C| = k$ such that every pair of vertices in $C$ is connected by an edge in $E$ ($C$ is a complete subgraph of $G$).
\end{quote}

In Clique problem, we want to check whether there exists a subset $C \subseteq V$ with $|C| = k$ such that:
\begin{equation}
    \forall u, v \in C, \quad (u,v) \in E.
\end{equation}

To reduce the Clique problem to SAT, we encode it as a Boolean formula in CNF. Let $n=|V|$ denote the number of vertices, the Boolean variables are defined as follows:
\begin{equation}
    x_{i,j}, \quad 1 \leq i \leq k, 1 \leq j \leq n,
\end{equation}
where $x_{i,j}=1$ if and only if vertex $j$ is in the $i$-th position of the clique. Otherwise, $x_{i,j} = 0$.
We introduce the following constraints for Clique:

\textbf{Constraint I}: each clique position is occupied by at least one vertex:
    \begin{equation}
        \bigvee_{j=1}^{n} x_{i,j}, \quad \forall i \in \{1, \dots, k\}.
    \end{equation}

\textbf{Constraint II}: each vertex appears at most once in the clique:
    \begin{equation}
        \neg x_{i,j} \vee \neg x_{i',j}, \quad \forall i \neq i', \forall j.
    \end{equation}

\textbf{Constraint III}: each clique position contains a different vertex:
    \begin{equation}
        \neg x_{i,j} \vee \neg x_{i,j'}, \quad \forall j \neq j', \forall i.
    \end{equation}

\textbf{Constraint IV}: selected vertices must be connected in $G$. For every non-adjacent pair $(u, v) \notin E$:
    \begin{equation}
        \neg x_{i,u} \vee \neg x_{i',v}, \quad \forall i \neq i', \text{ where } (u, v) \notin E.
    \end{equation}

The conjunction of all these constraints forms a CNF formula that is satisfiable if and only if there exists a clique of size $k$ in $G$.
Let $|E|$ denote the number of edges in $G$. In calculating $\#v$ and $\#c$ of 3-SAT optimizer, we have assumed $n>3$ to avoid including $\min(n-3,1)$ in the final numbers. This is reasonable since otherwise the problem is too easy. The number of clauses, variables and length of clauses of Clique are showin in Table.~\ref{tab:v_c_clique}.

\begin{table}[h]
\centering
\caption{Number of variables ($\#v$), clauses ($\#c$), average ($l_{avg}$) and total ($l_{total}$) length of mixed k-SAT and 3-SAT optimizer with our formulation of Clique.}
\renewcommand{\arraystretch}{1.5} 
\begin{tabular}{c|cc}
\hline
Optimizer & Mixed k-SAT & 3-SAT\\
\hline
$\#v$ & $kn$ & $\frac{k^2+k}{4}n^2+\frac{k^2+5k}{4}n-(3k+\frac{k(k-1)}{2}|E|)$ \\
$\#c$ & $\frac{k^2+k}{4}n^2+\frac{3k^2-k}{4}n+k-\frac{k(k-1)}{2}|E| $ & $\frac{k^2+k}{2}n^2+k^2n-[2+(k-1)|E|]k$  \\
$l_{avg}$ & $\frac{\#c}{l_{total}}$ & $3$ \\
$l_{total}$ & $\frac{k^2+k}{2}n^2 + \frac{k^2-k}{2}n-k(k-1)|E|$ & $3\frac{k^2+k}{2}n^2+3k^2n-3[2+(k-1)|E|]k$ \\
\hline
\end{tabular}
\label{tab:v_c_clique}
\end{table}

\subsection{Set Packing}
The definition of Set Packing is:
\begin{quote}
  Let $U = \{u_1, u_2, \ldots, u_n\}$ be a finite universe, and let 
  $\mathcal{S} = \{S_1, S_2, \ldots, S_m\}$ be a family of subsets where each $S_i \subseteq U$. 
  Given an integer $k$, is there a subfamily $\mathcal{S}' \subseteq \mathcal{S}$ of size $k$ such that 
  any two distinct sets $S_i, S_j \in \mathcal{S}'$ satisfy $S_i \cap S_j = \varnothing$.
\end{quote}

Set Packing can be easily reduced to Clique by constructing a graph with node $v_i$ representing $S_i$. There exists a edge $(i,j)$ if and only if $S_i \cap S_j = \varnothing$. Thus Set Packing can be reduced to Clique, then SAT and Graph Coloring, and finally optimized by our oscillators.

\subsection{Node Cover}
The Node (Vertex) Cover problem is defined as follows:
\begin{quote}
    Given a directed graph $G = (V, E) $ and an integer $k$, does there exist a subset of vertices $C \subseteq V$ with $|C| = k$ such that every directed edge in $E$ has at least one endpoint in $C$?
\end{quote}

Node Cover can be reduced to Clique: graph $G$ has a Node Cover with $k$ vertices if and only if $\overline{G}$ has a Clique of size $|V| - k$, where $\overline{G}$ is the complement of $G$ and $V$ is the set of vertices.
We can also reduce Node Cover directly to SAT by introducing Boolean variables $x_{v,i}$:
\begin{equation}
x_{v,i}, \quad \forall v \in V, \quad \forall i \in \{1, 2, \dots, k\}
\end{equation}

where $x_{v,i} = 1$ if vertex $v$ is the $i$-th vertex in the cover. Otherwise, $x_{v,i} = 0$.
The following constraints ensure that the selected set forms a valid Node Cover of size $k$.

\textbf{Constraint I}: each position in the cover must be occupied by at least one vertex:
    \begin{equation}
    \bigvee_{v \in V} x_{v,i}, \quad \forall i \in \{1, \dots, k\}
    \end{equation}
    
\textbf{Constraint II}: No vertex appears in multiple positions:
    \begin{equation}
    \neg x_{v,i} \vee \neg x_{v,j}, \quad \forall v \in V, \quad \forall i \neq j
    \end{equation}

\textbf{Constraint III}: each position in the cover should contain exactly one vertex:
    \begin{equation}
    \neg x_{u,i} \vee \neg x_{v,i}, \quad \forall u \neq v, \quad \forall i \in \{1, \dots, k\}
    \end{equation}

\textbf{Constraint IV}: every edge $(u, v) \in E$ must be covered by at least one vertex:
    \begin{equation}
    (x_{u,1} \vee x_{v,1}) \vee (x_{u,2} \vee x_{v,2}) \vee \dots \vee (x_{u,k} \vee x_{v,k}), \forall (u,v) \in E
    \end{equation}

The above constraints are converted into CNF form, making them suitable for our Clause network. 
If the formula is satisfiable, the satisfying assignment corresponds to a valid Vertex Cover of size $k$. If unsatisfiable, no such cover exists.
The number of variables and clauses of Node Cover is shown in Table.~\ref{tab:v_c_node_cover}.

\begin{table}[h]
\centering
\caption{Number of variables ($\#v$), clauses ($\#c$), average ($l_{avg}$) and total ($l_{total}$) length of mixed k-SAT and 3-SAT optimizer with our formulation of Node Cover.}
\setlength{\tabcolsep}{8pt} 
\renewcommand{\arraystretch}{1.5} 
\begin{tabular}{c|cc}
\hline
Optimizer & Mixed k-SAT & 3-SAT\\
\hline
$\#v$ & $k|V|$ & $-3k+|E|(2k-3)+\frac{1}{2}k|V|(k+|V|+2)$ \\
$\#c$ & $k+|E|+\frac{1}{2}k|V|(k+|V|-2)$ & $-2k+2|E|(k-1)+k|V|(k+|V|-1)$  \\
$l_{avg}$ & $\frac{k|V|+2k|E|+k|V|(k+|V|-2)}{k+|E|+\frac{1}{2}k|V|(k+|V|-2)}$& $3$ \\
$l_{total}$ & $k|V|+2k|E|+k|V|(k+|V|-2)$ & $-6k+6|E|(k-1)+3k|V|(k+|V|-1)$ \\
\hline
\end{tabular}
\label{tab:v_c_node_cover}
\end{table}

\subsection{Set Covering}
Problem definition:
\begin{quote}
    Given a finite set $\mathcal{U} = \{u_1, u_2, \dots, u_m\}$, and a collection of subsets $\mathcal{S} = \{S_1, S_2, \dots, S_n\}$ such that each $S_j \subseteq \mathcal{U}$, the goal of Set Covering is to decide whether there is a sub-collection $\mathcal{C} \subseteq \mathcal{S}$ such that $\bigcup_{S \in \mathcal{C}} S = \mathcal{U}$ (i.e., every element in $\mathcal{U}$ is contained in at least one set in $\mathcal{C}$), and $|C| = k$, where $k$ is an integer.
\end{quote}

We define binary variable $x_i=1$ if and only if $S_i$ is selected in $\mathcal{C}$.
We address the optimization version of Set Cover, which belongs to NP-hard:
\begin{equation}
\begin{aligned}
\text{min} \quad & \sum_{j=1}^{n} x_j \\
\text{s.t.} \quad & \sum_{j : u_i \in S_j} x_j \geq 1, \quad \forall u_i \in \mathcal{U} \\
& x_j \in \{0,1\}, \quad \forall j = 1, \ldots, n
\end{aligned}
\label{eq:set_covering}
\end{equation}

$j : u_i \in S_j$ means the set of $S_j$ that contains $u_i$.
The objective function can be implemented by Weighted Sum network.
The constraint $\sum_{j : u_i \in S_j} x_j \geq 1$ in Eq.~\ref{eq:set_covering} can be converted to CNF:
\begin{equation}
    (\sum_{j : u_i \in S_j} x_j \geq 1) \iff (\bigvee_{j : u_i \in S_j} x_{j})
    \label{eq:geq1}
\end{equation}

\subsection{Chromatic Number}
\begin{quote}
    Let $G=(V,E)$ be a finite, simple, undirected graph.  A proper $k$‑coloring of $G$ is a mapping
  $\;c:V \to \{1,2,\dots,k\}$ such that
  \begin{equation}
      \forall\,(u,v)\in E \;:\; c(u) \neq c(v).
  \end{equation}
  The chromatic number of $G$, denoted $\chi(G)$, is the
  smallest integer $k$ for which a proper $k$-colouring exists:
  \begin{equation}
      \chi(G) \;=\; \min \bigl\{ k \in \mathbb{N} \;\bigm|\;
      \text{$G$ admits a proper $k$‑colouring}\bigr\}.
  \end{equation}
\end{quote}

The Chromatic Number problem can be reduced to a series of successive n-Graph Coloring tests. 
4 and 3-Graph Coloring can directly approximated by oscillators~\cite{richelle_graph_coloring}. 
For $n$-Graph Coloring with $n\geq5$, we define $x_{v,i}=1$ if node $v$ is colored with color $i$. Otherwise, $x_{v,i}=0$. Let $\mathcal{N}$ be the set of neighboring nodes, and $V$ is the set of nodes. The overall optimization problem is:
\begin{equation}
\begin{aligned}
\text{min} \quad & \sum_{(uv) \in E} \sum_{i=1}^{n} x_{u,i} x_{v,i} \\
\text{s.t.} \quad & \sum_{i=1}^{k} x_{v,i} = 1, \quad \forall v \in \mathcal{V} \\
\end{aligned}
\label{eq:obj_chromatic}
\end{equation}

The objective function can be implemented by Energy AND network.
$\sum_{i=1}^{k} x_{v,i} = 1$ in Eq.~\ref{eq:obj_chromatic} can be converted to CNF: 
\begin{equation}
    (\sum_{i=1}^{k} x_{v,i} = 1) \iff ((\bigvee_{1\leq i \leq k} x_{v,i}) \wedge (\bigwedge_{1\leq i < j \leq k} (\neg x_{v,i} \vee \neg x_{v,j}))
    \label{eq:CNF_equal_one}
\end{equation}

The objective function is implemented by Energy AND network. 
For the Clause network of Chromatic Number, the numbers of variables and clauses are shown in Table.~\ref{tab:v_c_chromatic}

\begin{table}[h]
\centering
\caption{Number of variables ($\#v$), clauses ($\#c$), average ($l_{avg}$) and total ($l_{total}$) length of mixed k-SAT and 3-SAT optimizer with our formulation of Clause network in Chromatic Number.}
\setlength{\tabcolsep}{8pt} 
\renewcommand{\arraystretch}{1.5} 
\begin{tabular}{c|cc}
\hline
Optimizer & Mixed k-SAT & 3-SAT\\
\hline
$\#v$ & $k|V|$ & $k|V|+\frac{1}{2}k(k+1)-3$\\
$\#c$ & $1+\frac{1}{2}k(k-1)$ & $k^2-2$ \\
$l_{avg}$ & $\frac{k^2}{1+\frac{1}{2}k(k-1)}$& $3$ \\
$l_{total}$ & $k^2$ & $3k^2-6$ \\
\hline
\end{tabular}
\label{tab:v_c_chromatic}
\end{table}

\subsection{Feedback Node Set}

The NP-hard version of Feedback Node Set problem is:
\begin{quote}
    Given a directed graph $H = (V, E)$ and a positive integer $k$, find the smallest size of the subset of vertices (nodes) $R \subseteq V$ such that every directed cycle in $H$ contains at least one vertex from $R$.
\end{quote}

We follow Karp's definition~\cite{karp} which focuses on the directed graph version.
Proved by~\cite{ising_many_np}, a directed graph $G$ is acyclic if and only if we can assign each vertex $v$ with a height $h(v)$, so that for each $(u,v)\in E$, $h(u)<h(v)$.
We set binary variable $x_{v,i}=1$ if and only if $v$ is assigned height $i$, $1\leq i \leq |V|$.
For each $v$, $\sum_ix_{v,i}=0$ if and only if $x_{v,i}=0$. This means that $v$ belongs to the Feedback Vertex set and can not be assigned a valid height.
$\sum_ix_{v,i}=1$ if there exist a $x_{v,i}=1$ for this $v$: $v$ is assigned height $i$, $h(v)=i$, and $v$ is in a directed circle. 
$\sum_ix_{v,i}$ is also constrained to be less or equal to 1.
To constrain that an edge only connects a node with lower height to a node at higher height, and to minimize the feedback set, we set overall Potts Hamiltonian as:
\begin{equation}
\label{fns}
\begin{aligned}
\text{min} \quad & \sum_{(u,v)\in E} \sum_{i\geq j} x_{u,i}x_{v,j} + \lambda \sum_{v} (1-\sum_ix_{v,i})\\
\text{s.t.} \quad & \sum_ix_{v,i} \leq 1,
\end{aligned}
\end{equation}

where $\lambda$ is between $0$ and $1$. $\sum_{(u,v)\in E} \sum_{i\geq j} x_{u,i}x_{v,j}$ is implemented by Energy AND network. $\sum_{v} (1-\sum_ix_{v,i})$ is implemented by the Weighted Sum network.
The constraint can be converted to CNF:
\begin{equation}
    (\sum_ix_{v,i} \leq 1) \iff \bigwedge_{1\leq i<j \leq |V|} (\neg x_{v,i} \vee \neg x_{v,j})
\end{equation}

For the Clause network of Feedback Node Set, the numbers of variables and clauses are shown in Table.~\ref{tab:v_c_feedback_node_set}

\begin{table}[h]
\centering
\caption{Number of variables ($\#v$), clauses ($\#c$), average ($l_{avg}$) and total ($l_{total}$) length of mixed k-SAT and 3-SAT optimizer with our formulation of Clause network in Feedback Node Set.}
\setlength{\tabcolsep}{8pt} 
\renewcommand{\arraystretch}{1.5} 
\begin{tabular}{c|cc}
\hline
Optimizer & Mixed k-SAT & 3-SAT\\
\hline
$\#v$ & $|V|^2$ & $\frac{1}{2}|V|(3|V|-1)$\\
$\#c$ & $\frac{1}{2}|V|(|V|-1)$ & $|V|(|V|-1)$ \\
$l_{avg}$ & $2$& $3$ \\
$l_{total}$ & $|V|(|V|-1)$ & $3|V|(|V|-1)$ \\
\hline
\end{tabular}
\label{tab:v_c_feedback_node_set}
\end{table}


\subsection{Feedback Arc Set}

The NP-hard version of Feedback Arc Set problem for directed graphs is defined as:
\begin{quote}
    Given a directed graph $H = (V, E)$ and a positive integer $k$, find the smallest subset of edges $F \subseteq E$ such that every directed cycle in $H$ contains at least one edge from $F$.
\end{quote}

We assign a binary variable $y_{uv} = 1$ for each edge $(u,v)\in E$, indicating that the edge is retained in the graph, and is not part of the Feedback Arc Set. $y_{uv} = 0$ means the edge is removed from the graph.
To ensure acyclicity, we assign each vertex $v$ a height via binary variables $x_{v,i}$, where $x_{v,i} = 1$ if and only if node $v$ is assigned height $i$. 
We also define $x_{uv,i}=1$ if both $y_{uv}=1$ and $x_{u,i}=1$~\cite{ising_many_np}.
Each vertex must be assigned a unique height, enforced by:
\begin{equation}
\label{fas_1}
\sum_i x_{v,i} = 1, \quad \forall v \in V    
\end{equation}

Similar to Feedback Node Set, we require that any retained edge $(u,v)$ must go from lower to higher height, i.e., $h(u) < h(v)$. 
This is expressed via:
\begin{equation}
\label{fas_2}
y_{uv} = \sum_{i} x_{uv,i}, \quad \forall uv \in E
\end{equation}

If $x_{uv,i}=1$, then $u$ is assigned height $i$:
\begin{equation}
    \text{if} \quad x_{uv,i}=1, \quad \text{then} \quad x_{u,i} = 1
\end{equation}

If $x_{uv,i}=1$, then $(uv)$ is not removed from the graph, and it is required that a edge points from lower to higher height:
\begin{equation}
    \text{if} \quad x_{uv,i}=1, \quad \text{then} \quad \sum_{j>i} x_{v,j}=1
\end{equation}

We set the objective function to minimize the number of removed edges. 
The overall Potts Hamiltonian is:
\begin{equation}
\begin{aligned}
\text{min} \quad & \sum_{uv} (1 - y_{uv}) \\
\text{s.t.} \quad & \sum_i x_{v,i} = 1, \quad \forall v\in V \\ 
& y_{uv} = \sum_{i} x_{uv,i}, \quad \forall uv \in E \\
& \text{if} \quad x_{uv,i}=1, \quad \text{then} \quad x_{u,i} = \sum_{j>i} x_{v,j}=1, \quad \forall uv \in E \quad \forall i
\end{aligned}
\end{equation}

The if-then can be transformed to energy minimization, and $y_{uv}$ can be replaced by $\sum_{i} x_{uv,i}$, so the Potts Hamiltonian becomes:
\begin{equation}
\begin{aligned}
\text{min} \quad & \lambda \sum_{uv} (1 - \sum_{i} x_{uv,i}) + \sum_{uv} \sum_{i} x_{uv,i} (1-x_{u,i}) + \sum_{uv} \sum_{i} x_{uv,i} (1-\sum_{j>i} x_{v,j}) \\ 
\text{s.t.} \quad & \sum_i x_{v,i} = 1, \quad \forall v\in V \\ 
\end{aligned}
\end{equation}

$0<\lambda<1$ makes sure that the constraint is satisfied before minimizing Feedback Arc Set. The objective function can be further transformed to fit for Energy AND and Weighted Sum network:
\begin{equation}
\begin{aligned}
\text{min} \quad & (2-\lambda) \sum_{uv} \sum_{i} x_{uv,i} - \sum_{uv} \sum_{i} x_{uv,i} x_{u,i} - \sum_{uv} \sum_{i} \sum_{j>i} x_{uv,i} x_{v,j} \\ 
\text{s.t.} \quad & \sum_i x_{v,i} = 1, \quad \forall v\in V \\ 
\end{aligned}
\end{equation}

The constrain can be transformed to CNF similar to Eq.~\ref{eq:CNF_equal_one}. The number of variables and clauses for the constraint is similar to Table.~\ref{tab:v_c_chromatic}.


\subsection{Clique Cover}

\begin{quote}
Given a graph $G = (V, E)$ and integer $k$, determine whether there exists a partition $V = V_1 \cup V_2 \cup \dots \cup V_k$, such that for all $1 \leq i \leq k$, the subgraph $G[V_i]$ is a clique.
\end{quote}

The Clique Cover problem on a graph $G$ is equivalent to the Graph Coloring problem on its complement graph $\overline{G}$. Complement graph $\overline{G} = (V, \overline{E})$ is defined such that $(u,v) \in \overline{E}$ if and only if $u \neq v$ and $(u,v) \notin E$.
A $k$-clique cover in $G$ corresponds to a proper $k$-coloring in $\overline{G}$.
In $\overline{G}$, a proper coloring ensures that no two adjacent vertices have the same color. This implies that in the original graph $G$, all vertices of the same color class are pairwise connected, i.e., form a clique.
Therefore, Clique Cover can be trivially reduced to Graph Coloring and approximated by oscillators.

\subsection{Exact Cover}
\begin{quote}
      Let $U=\{u_1,u_2,\dots,u_n\}$ be a finite ground set and
  $S=\{S_1,S_2,\dots,S_m\}$ a family of subsets of $U$. $\bigcup_{i} S_i \;=\; U$.
  A Exact Cover of $U$ is a sub‑family $C\subseteq S$ such that:
  \begin{equation}
      \bigcup_{S_i\in C} S_i \;=\; U
      \quad\text{and}\quad
      S_i\cap S_j=\varnothing \; \text{ for all } S_i\neq S_j\in C.
  \end{equation}
  Exact Cover is the decision problem that asks whether an
  exact cover exists for the given $(U,S)$.
\end{quote}

For every set $S_i\in S$, we introduce one Boolean variable $x_i$. $x_i=1$ if and only if $S_i\in C$.
Fix an element $u_j\in U$ and let $\operatorname{idx}(u_j)=\{\,i\mid u_j\in S_i\}$.
For example, if $u_1$ only belongs to $S_2$ and $S_5$, then $\operatorname{idx}(u_j)=\{2,5\}$.
We need at least one chosen set and no two chosen sets overlap on $u_j$. Exact Cover can be reduced to SAT:
\begin{equation}
(\bigvee_{i\in\operatorname{idx}(u_j)} x_i) \land (\bigwedge_{\substack{i<\ell\\ i,\ell\in\operatorname{idx}(u_j)}} \neg x_i \lor \neg x_\ell), \quad \forall j = 1,\text{...},n
\end{equation}

Let $s_j$ denote the size of $\operatorname{idx}(u_j)$, the numbers of variables and clauses in Exact Cover are shown in Table.~\ref{tab:v_c_exact_cover}

\begin{table}[h]
\centering
\caption{Number of variables ($\#v$), clauses ($\#c$), average ($l_{avg}$) and total ($l_{total}$) length of mixed k-SAT and 3-SAT optimizer with our formulation of Exact Cover. $s_j=|\operatorname{idx}(u_j)|$.}
\setlength{\tabcolsep}{8pt} 
\renewcommand{\arraystretch}{1.5} 
\begin{tabular}{c|cc}
\hline
Optimizer & Mixed k-SAT & 3-SAT\\
\hline
$\#v$ & $m$ & $\sum_{j=1}^n \frac{1}{2}s_j (s_j+1) - 3n$\\
$\#c$ & $\sum_{j=1}^n \frac{1}{2}s_j (s_j+1)$ & $\sum_{j=1}^n s_j(s_j+1) - 3n$ \\
$l_{avg}$ & $\frac{2\sum_{j=1}^n s_j^2}{\sum_{j=1}^n s_j (s_j+1)}$& $3$ \\
$l_{total}$ & $\sum_{j=1}^n s_j^2$ & $3\sum_{j=1}^n s_j(s_j+1) - 9n$ \\
\hline
\end{tabular}
\label{tab:v_c_exact_cover}
\end{table}


\subsection{Hitting Set}
Problem definition:
\begin{quote}
    Given a collection of subsets $\mathcal{C} = \{C_1, C_2, \dots, C_n\}$ over a universe $\mathcal{U} = \bigcup_{i=1}^n C_i$, and an integer $k$,, the goal of Hitting Set is to decide whether there exists a subset $H \subseteq \mathcal{U}$ of size at most $k$ such that $H \cap C_i \neq \emptyset$ for every $C_i \in \mathcal{C}$.
\end{quote}

We can reduce the Hitting Set problem to the Set Covering problem by exchanging the roles of elements and sets:

\begin{itemize}
    \item Construct a new universe $\mathcal{U}' = \{C_1, C_2, \dots, C_n\}$ consisting of the original subsets as elements.
    \item For each element $e_j \in \mathcal{U}$, define a subset $S_j = \{C_i \in \mathcal{C} \mid e_j \in C_i\}$. For example, if $e_j$ only belongs to $C_1$ and $C_5$, then $S_j=\{C_1,C_5\}$. 
    \item The new collection of subsets is $\mathcal{S} = \{S_j \mid e_j \in \mathcal{U}\}$.
\end{itemize}

Then, selecting a subset $\mathcal{C}' \subseteq \mathcal{S}$ of size at most $k$ that covers $\mathcal{U}'$ (Set Covering) is equivalent to selecting a hitting set $H \subseteq \mathcal{U}$ of size at most $k$ that hits all sets in $\mathcal{C}$.



\subsection{3-Dimensional Matching}

The problem is defined as follows:
\begin{quote}
Given a finite set $T$ and a collection $U \subseteq T \times T \times T$ of triples, does there exist a subset $W \subseteq U$ such that $|W| = |T|$ and no two elements of $W$ agree in any coordinate? That is, for any two distinct triples $(a,b,c), (a',b',c') \in W$, we have $a \ne a'$, $b \ne b'$, and $c \ne c'$.
\end{quote}

We reduce the 3-Dimensional Matching problem to an instance of the Set Packing problem.
Given an instance of 3DM with a finite set $T$ and a collection $U \subseteq T \times T \times T$, we construct the Set Packing instance as follows:
\begin{itemize}
    \item Finite universe $U = T^{(1)} \cup T^{(2)} \cup T^{(3)}$, where $T^{(1)}$, $T^{(2)}$, and $T^{(3)}$ are three disjoint copies of $T$ representing the first, second, and third coordinates respectively.
    \item For each triple $(a, b, c) \in U$, define a subset $S_{(a,b,c)} = \{a^{(1)}, b^{(2)}, c^{(3)}\} \subseteq X$.
    \item Let $\mathcal{S} = \{S_{(a,b,c)} \mid (a,b,c) \in U\}$ and set $k = |T|$.
\end{itemize}

Then, a solution to the Set Packing instance with $k$ disjoint subsets corresponds to a valid matching $W \subseteq U$ of size $|T|$ where no two triples share a coordinate.

\subsection{Number Partitioning}
The Number Partitioning problem in this paper corresponds to Partitioning in Karp's 21 NP paper~\cite{karp}.
Given a finite set of positive integers:
\begin{equation}
    S = \{s_1, s_2, \dots, s_n\}, \quad s_i \in \mathbb{Z}^+
\end{equation}

The goal of Number Partitioning is to partition the set $S$ into two disjoint subsets $S_1$ and $S_2$ such that:
\begin{equation}
S_1 \cup S_2 = S, \quad S_1 \cap S_2 = \emptyset
\end{equation}

and the absolute difference between the sums of the two subsets is 0:
\begin{equation}
 \left| \sum_{s \in S_1} s - \sum_{s \in S_2} s \right| = 0
\end{equation}

Equivalently, define binary variables $x_i \in \{0,1\}$, where:
\begin{equation}
x_i = \begin{cases}
1 & \text{if } s_i \in S_1 \\
0 & \text{if } s_i \in S_2
\end{cases}
\end{equation}

Then the problem becomes:
\begin{equation}
    \sum_{i=1}^{n} s_i x_i  = \frac{1}{2} \sum_{i=1}^{n} s_i.
\end{equation}

We can formulate an 0-1 Integer Programming (IP) for Number Partitioning. Let $b = \frac{1}{2} \sum_{i=1}^{n} s_i$, denoting half of the full price. The IP is defined as:
\begin{equation}
    \sum_{i=1}^{n} s_ix_i = b.
\end{equation}

\subsection{Knapsack}
The definition of the decision version of Knapsack, which is an NP-complete problem, is:
\begin{quote}
Given integers $(a_1, a_2, \dots, a_r, b) \in \mathbb{Z}^{r+1}$, the goal of the Knapsack problem is to determine whether there exists a binary vector $x = (x_1, \dots, x_r) \in \{0,1\}^r$ such that:
\begin{equation}
    \sum_{j=1}^{r} a_j x_j = b.
\end{equation}
\end{quote}

The problem is identical to 0-1 Integer Programming in Eq.~\ref{eq:ip}.

\subsection{Graph Partitioning and Min Cut}
Graph Partitioning is a NP-complete problem, and Min Cut is a special case of it, which is P. Graph Partitioning is defined as:
\begin{quote}
    Given an undirected graph $ G = (V, E) $, divide the vertex set $ V $ into two disjoint subsets $ V_1 $ and $ V_2 $ such that:
\begin{itemize}
  \item $ V_1 \cup V_2 = V $ and $ V_1 \cap V_2 = \emptyset $,
  \item The sizes of the two subsets are balanced, i.e., $ \left| |V_1| - |V_2| \right| =0 $, where we assume that  $ n = \left| |V|\right| $ is an even number,
  \item The number of edges that have one endpoint in $ V_1 $ and the other in $ V_2 $ (called cut edges) is minimized:
  \begin{equation}
    \text{Cut}(V_1, V_2) = \left| \{ (u, v) \in E \mid u \in V_1, v \in V_2 \} \right|
  \end{equation}
\end{itemize}
\end{quote}

We represent each edge with an oscillator.
First, we use a large coupling strength $J_1 >> 1$ to fully connect the oscillators ($\frac{1}{2} n(n-1)$ connections in total) to satisfy the $ \left| |V_1| - |V_2| \right| =0 $ constraint, since oscillators will minimize Potts Hamiltonian: $\sum_{i,j\in E} cos(\theta_i - \theta_j) J_1$.
Then, we use $||J_2|| << J_1$ to find an optimal solution, without violating the constraint. $J_2  = -1$ connects oscillator $i$ and $j$ if $(i,j)\in E$. Else, $J_2 = 0$. Let the number of edges between $||V_1||$ and $||V_2||$ be $m$.
This will minimize the Potts Hamiltonian:
\begin{equation}    
\sum_{(i,j)\in E, i\in V_1, j\in V_2} cos(180) J_2 + \sum_{(i,j)\in E, i,j\in V_1} cos(0) J_2 + \sum_{(i,j)\in E, i,j\in V_2} cos(0) J_2.
\end{equation}

The overall coupling strength between each oscillator is $J_1 + J_2$.
Graph Partitioning is Min Cut if we ignore the $\left| |V_1| - |V_2| \right| =0$ constraint.

\subsection{Independent Set}
The Independent Set is defined as:
\begin{quote}
  Let $G = (V, E)$ be an undirected graph. A subset of vertices
  $S \subseteq V$ is called an independent set if no two vertices
  in $S$ are adjacent to each other in $G$. Formally, for every pair of
  vertices $u, v \in S$, it must hold that $\{u,v\} \notin E$. 
  The Independent Set problem asks, given a graph $G$ and an integer
  $k$, whether there exists an independent set $S$ of size at least $k$.
\end{quote}

If $S$ is an independent set in $G$, then no two vertices in $S$ can share an edge. Therefore, if $S$ is an independent set, $V \setminus S$ must be a Node Cover, because for any edge
$\{u,v\} \in E$, at least one of $u$ or $v$ must be outside $S$.

\subsection{Max-Cut}
Max-Cut is defined as follows:

\begin{quote}
    Given an undirected graph $G = (V,E)$, a weighting function $w : E \to \mathbb{Z}$, and a positive integer $W$, determine whether there exists a subset $S \subseteq V$ such that the total weight of the cut edges is at least $W$:
    \begin{equation}
        \sum_{\substack{(u,v) \in E \\ u \in S, v \notin S}} w(u,v) \;\;\geq\; W.
    \end{equation}
\end{quote}

We associate a binary variable $x_i \in \{0,1\}$ with each vertex $v_i \in V$. If $x_i = 1$, vertex $v_i$ is placed in subset $S$; if $x_i = 0$, then $v_i$ is placed in the complement $V \setminus S$. For each edge $(u,v) \in E$, the indicator of whether the edge is cut can be expressed as the XOR of the vertex variables:
\begin{equation}
    y_{uv} = x_u \oplus x_v = |x_u - x_v|.
\end{equation}

Thus, the cut weight is
\begin{equation}
    \text{Cut}(x) \;=\; \sum_{(u,v)\in E} w(u,v)\,|x_u - x_v|.
\end{equation}

This can be reduced to an energy minimization form. Since $x_i \in \{0,1\}$, $|x_u - x_v| = x_u + x_v - 2x_u x_v$.
The Potts Hamiltonian becomes:
\begin{equation}
    \text{min} \quad \text{Cut}(x) \;=\; \sum_{(u,v)\in E} w(u,v)\,(x_u + x_v - 2x_u x_v).
\end{equation}

This can be implemented by Energy AND and Weighted Sum networks.

\section{More Applications of Oscillator Optimizer}
\subsection{Maze Solver}
\begin{figure}[h]
	\begin{center}
    \includegraphics[width=0.8\linewidth]{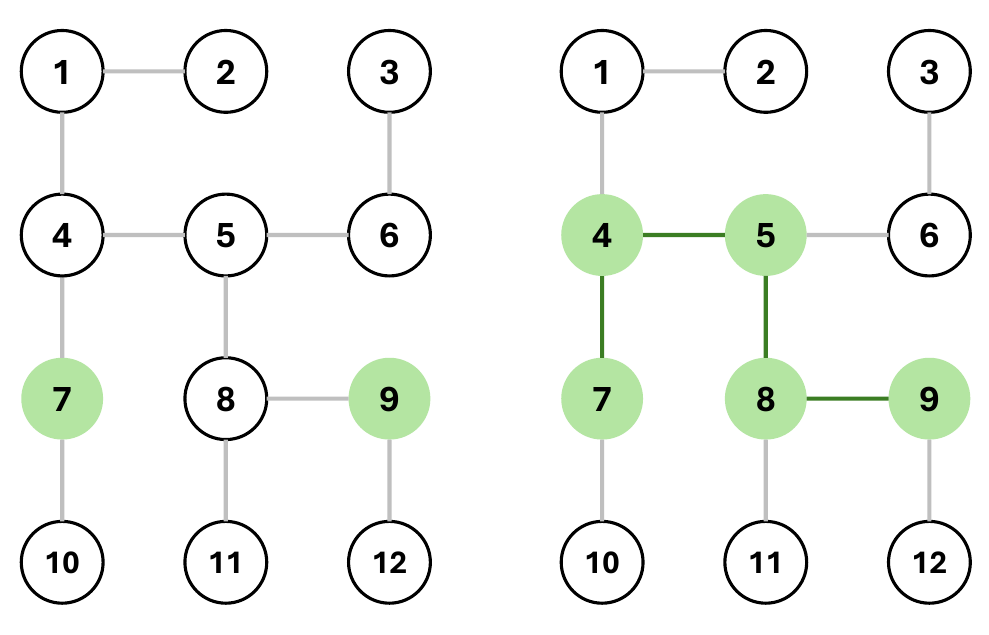}
	\end{center}
\caption{An example of maze problem. Left: each node is represented by an oscillator. 7 and 9 are fixed at True phase, indicating start and end nodes. Right: the solution of this maze problem.}
	\label{fig:maze_problem}
\end{figure}

\begin{figure}[h]
	\begin{center}
    \includegraphics[width=0.9\linewidth]{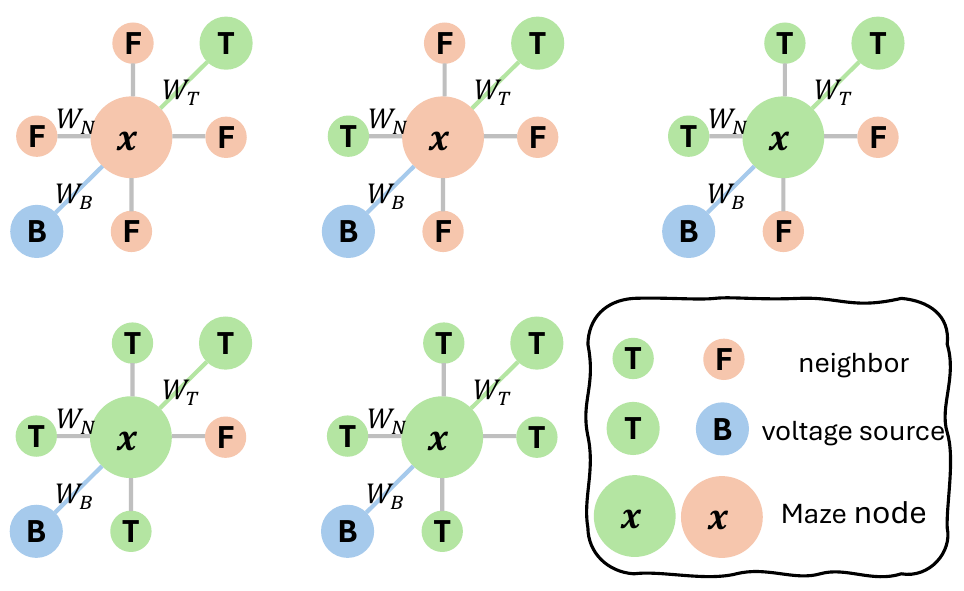}
	\end{center}
\caption{Truth table of oscillator maze solver. $x$ will converge to True if and only if it is connected to at least 2 True oscillators, which indicates that it may be on the solution path.}
	\label{fig:maze_truth_table}
\end{figure}

\begin{figure}[h]
	\begin{center}
    \includegraphics[width=1\linewidth]{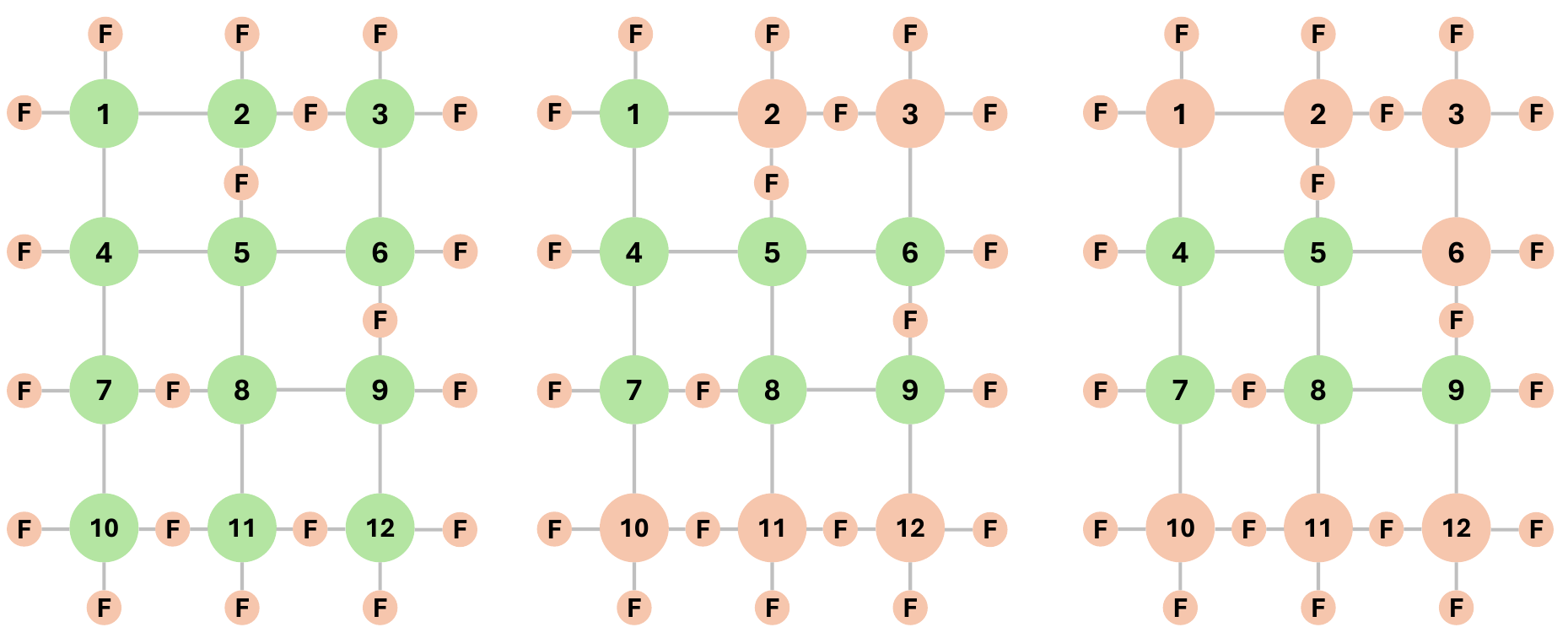}
	\end{center}
\caption{Phase update of a maze instance. The oscillators on the correct path converge to True phase, while other oscillators settle down in False phase.}
	\label{fig:maze_update}
\end{figure}

We assume that the maze has a unique solution, which also implies that no loops exist within it. 
An example instance is shown in Fig.~\ref{fig:maze_problem}, where the goal is to find the path between node 7 and node 9. After the oscillator phases converge, nodes on the maze solution path will converge to True phase and all other nodes will go to False phase.
At each node, there are four possible directions to go on the maze: up, down, left, and right. If a path exists in a given direction, the node is connected to another oscillator in that direction. If no path exists, it is instead connected to an auxiliary oscillator that is fixed in the False phase. A node with fewer than two accessible directions cannot be part of the correct path; during the phase updates, it will eventually switch to False. Conversely, if a node has two or more accessible directions, it may belong to the correct path and its phase is set to True. The corresponding truth table is shown in Fig.~\ref{fig:maze_truth_table}. 
$W_N$ is the connection between an oscillator and its neighbors, including auxiliary oscillators. Each node oscillator is also connected to the voltage source Blue and True, with connection weights $W_B$ and $W_T$. $W_B$ is set to a large value so the oscillator can only choose between True and False.
The truth table requires that:
\begin{equation}
    W_T + 4W_N < 0, \quad W_T + 3W_N < W_N, \quad W_T + 2W_N < 2W_N, \quad 3W_N < W_T + W_N.
\end{equation}
This can be simplified to:
\begin{equation}
    2W_N < W_T < 0.
\end{equation}
All oscillators are initialized in the True phase, and their dynamics evolve are illustrated in Fig.~\ref{fig:maze_update}.

\subsection{Shortest Path}
Shortest Path is a special case of Steiner Tree and belongs to P problems. 
It is defined as:

\begin{quote}
    Given a graph $G=(V,E)$, two vertices $s,t \in V$, and a positive integer $K$, determine whether there exists a path from $s$ to $t$ of length at most $K$.
\end{quote}

We encode this problem as a SAT formula using variables to represent the position of each vertex along the candidate path. Define the variable set
\begin{equation}
        X = \{x_{v,i} \mid v \in V,\; 1 \leq i \leq K \},
\end{equation}
where $x_{v,i}$ is a Boolean variable that is true if and only if vertex $v$ appears at position $i$ in the path.
The SAT formula is constructed from the following constraints:

\textbf{Constraint I}: if $x_{v,i}$ is true for some $i \geq 2$, then the previous position $i-1$ must contain at least one neighbor $w \in N(v)$ of $v$:
    \begin{equation}
                \bigwedge_{2 \leq i \leq K} \bigwedge_{v \in V} 
        \Big( x_{v,i}=1 \;\Rightarrow\; \sum_{w \in N(v)} x_{w,i-1} \geq 1).
    \end{equation}

\textbf{Constraint II}: The path must reach the target vertex $t$ within $K$ steps:
    \begin{equation}
        (\sum_{1 \leq i \leq K}x_{t,i}) \geq 1
    \end{equation}

\textbf{Constraint III}: At the first position, only the source vertex $s$ or its neighbors may appear: 
    \begin{equation}
        \lnot x_{v,1}, \forall v \in V \setminus N(s)
    \end{equation}

\subsection{Logic Gates and Sequential Computing}
\begin{figure}[h]
	\begin{center}
    \includegraphics[width=0.8\linewidth]{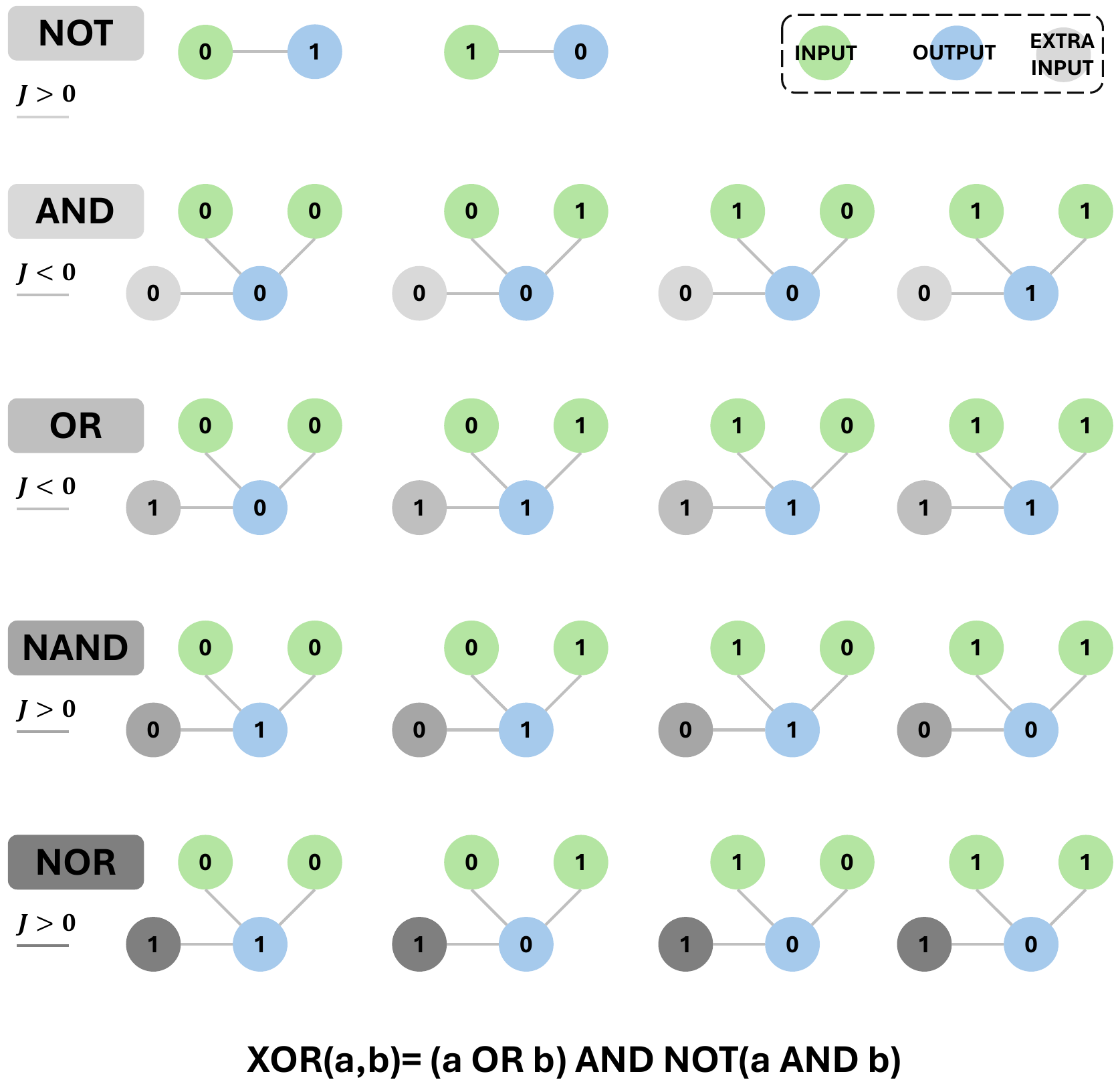}
	\end{center}
\caption{Logic gates implemented by 2-phase oscillators optimizing Potts Hamiltonian. The green input oscillators and gray additional oscillator specific to each logic gate, decide the phase of blue output oscillator like logic gates.}
	\label{fig:LogicGate}
\end{figure}

Fig.~\ref{fig:LogicGate} shows the design of logic gates with 2-phase oscillators.
The connection from the input and extra input to the output oscillator is unidirectional: the input affects the output, but the output does not influence the phase of the input.
They can form functionally complete set of logic gates so any logistic expressions can be represented by oscillators. Also, it can be calculated step-by-step: calculate a logic expression, then based on the result, calculate another one. 
However, XOR and XNOR can not be directly implemented by oscillators.

\subsection{Mixed k-SAT}
\begin{figure}[h]
	\begin{center}
    \includegraphics[width=0.8\linewidth]{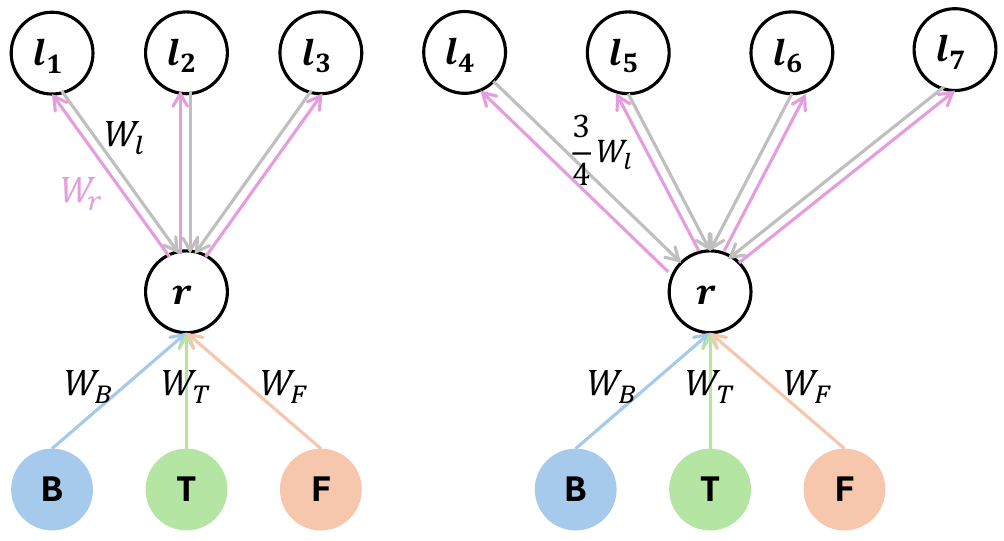}
	\end{center}
\caption{Extension of our optimizer to mixed k-SAT. In this figure we show an example SAT problem with 2 clauses, containing 3 and 4 variables each.}
	\label{fig:mixed_k_SAT}
\end{figure}

When reducing an NP problem to SAT, the length of each clause is not necessarily the same. If the optimizer only supports 3-SAT or a specific K-SAT, the reduction process requires additional redundant variables, placing a greater burden on the optimizer. Our Oscillator Optimizer can directly handle mixed K-SAT without extra problem transformation or the addition of variables. As shown in the Fig.~\ref{fig:mixed_k_SAT}, the weights can be scaled according to the length of each clause.

\subsection{Scaling with SAT Complexity}
\label{sec:osc_connect_sat}

\begin{table}[htbp] 
\centering
\caption{Number of nodes, connections, and maximum number of connections of a node (Max \# connections) in the oscillator optimizer with and without Biased Source. $v$ and $c$ are number of variables and clauses in SAT.}
\label{tab:num_node_connection}
\begin{tabular}{|c|ccc|}
\hline
\textbf{Designs} & \textbf{\# Nodes} & \textbf{\# Connections} & \textbf{Max \# connections} \\
\hline
Oscillator Optimizer     & $2v+c+3$ & $3v+6c+3$ & $\leq c+2$ \\
w/ Biased Source   & $2v+c+3$ & $3v+4c+3$ & $\leq c+2$ \\
\hline
\end{tabular}
\end{table}

As shown in Table.~\ref{tab:num_node_connection}, the number of oscillators and connections in our oscillator optimizer increases linearly with the complexity of the problem. 
The merging of True, False and Blue voltage sources into on Biased Source is especially helpful in reducing number of connections in large scale SAT problems. For example, in 110 variable 1400 clause SAT, the number of connections drops from 8733 to 5933.
It is important to note that since the hardware for the oscillator optimizer has not yet been designed, this analysis does not imply that the number of transistors and components in the hardware design would also increase linearly with the problem complexity.

\section{Conclusion}
In this paper, we formulate 19 of Karp's 21 NP-complete problems~\cite{karp}, and many other P problems, into oscillator formulations, enabling them to be addressed by our multi-phase oscillator optimizer. 
Oscillators represent a promising new computational fabric that departs from the traditional von Neumann paradigm, offering potential advantages in energy efficiency and scalability.



\section*{Acknowledgment}
This work was generously funded by the Stanford-Samsung Research Initiative, whose support is gratefully acknowledged.


\bibliographystyle{unsrt}
\bibliography{main}

\end{document}